\documentclass[conference]{IEEEtran}
\IEEEoverridecommandlockouts
\usepackage{cite}
\usepackage{amsmath,amssymb,amsfonts}
\usepackage{algorithm}
\usepackage{algorithmic}
\usepackage{graphicx}
\usepackage{subcaption}
\usepackage{textcomp}
\usepackage{xcolor}
\usepackage{endnotes}
\usepackage{url}
\usepackage{comment}
\newtheorem{proposition}{Proposition}
\def\BibTeX{{\rm B\kern-.05em{\sc i\kern-.025em b}\kern-.08em
		T\kern-.1667em\lower.7ex\hbox{E}\kern-.125emX}}
\begin{document}
	
	\title{CRB-Driven Beamforming and Trajectory Optimization for UAV-assisted ISAC System \\
	}
	
	\author{ 
		\IEEEauthorblockN{Yi Yang,  Qianqian Zhang, and Huaxia Wang
		}
		\IEEEauthorblockA{
			Department of Electrical and Computer Engineering, Rowan University, NJ, USA \\
			Emails: \url{{yangyi57, zhangqia, wanghu}}@rowan.edu}}	
	\maketitle
	
	\begin{abstract}
		In this paper, we study an unmanned aerial vehicle (UAV)-assisted integrated sensing and communication (ISAC) system, where a UAV enhances the sensing capability of a base station (BS) towards a target while ensuring reliable communication towards a downlink user. This architecture is practically attractive for future wireless networks due to the UAV's controllable mobility and adaptive sensing coverage in wireless environments.
		The sensing performance is characterized by the average Cramér–Rao bound (CRB), which quantifies the minimum variance of the unbiased angle-of-arrival estimation. To enhance the sensing performance, the UAV trajectory and beamforming parameters are jointly optimized under power and mobility constraints, while satisfying communication requirements to the downlink user. 
		To address the resulting non-convex problem, we employ null-space projection for beamforming design and adopt deep reinforcement learning for the trajectory optimization over a discrete-time scale. 
		In each time slot,  beamforming is optimized based on the channel state information to improve CRB performance while mitigating interference between the BS and the communication user. 
		Simulation results demonstrate that the proposed method significantly reduces the time-averaged CRB by over 10\%, compared with the ISAC system without UAV assistance, and also achieves a higher sensing accuracy than both the fixed-UAV-trajectory and the maximum-ratio-transmission-based beamforming benchmarks.
	\end{abstract}
	
	\begin{IEEEkeywords}
		integrated sensing and communication, unmanned aerial vehicle, deep reinforcement learning. %  Cramér–Rao bound (CRB), null-space projection, 
	\end{IEEEkeywords}
	\section{Introduction}
	The sixth-generation (6G) wireless network is expected to support a wide range of emerging applications, such as autonomous driving, smart home, and human activity sensing \cite{gonzalez2024integrated}. 
	Meeting these heterogeneous communication and sensing demands under limited spectrum and hardware resources, however, poses significant challenges \cite{liu2022integrated}. 
	Integrated sensing and communication (ISAC) technology has emerged as a promising 6G technology, where the dual-functional node of ISAC transmits a unified waveform to simultaneously perform sensing and communication on a shared hardware platform, thereby improving spectrum, energy, and hardware efficiency \cite{dong2024sensing}. 
	
	To improve the system performance, artificial noise (AN) has been widely incorporated into ISAC systems \cite{zou2024securing}. 
	Beyond its traditional role in enhancing physical-layer security, recent researches have shown that AN can also support sensing. 
	In particular, with appropriate beamforming design at the base station (BS), AN can degrade the interception capability of eavesdroppers while facilitating high-resolution sensing performance. 
	In \cite{hou2023secure}, the authors investigated the joint transmit beamforming of information and AN to maximize the communication secrecy rate under sensing constraints. 
	In \cite{lv2025isac}, a two-stage secure communication framework was proposed to use AN for sensing in the first stage and for jamming in the second stage, with the secrecy rate maximized via joint optimization of time allocation and beamforming designs. 
	However, in most existing works, AN competes with information signals for the limited transmit power at the BS, which restricts the sensing performance. To provide additional power sources in the ISAC system, unmanned aerial vehicles (UAVs) have attracted increasing interest due to their independent energy source, as well as the controllable mobility and flexible deployment \cite{huang2023unmanned}.
	Benefiting from their high altitudes, UAVs often maintain strong line-of-sight (LoS) links for reliable sensing and communication services. 
	In \cite{lyu2022joint}, the joint transmit beamforming and UAV trajectory design was investigated to maximize the average rate under sensing beampattern constraints. This line of work was extended in \cite{xiu2024improving} by incorporating secrecy rate considerations, and further in \cite{xiu2025secure} by introducing an intelligent reflecting surface (IRS) with joint optimization of UAV trajectory, transmit beamforming, and IRS phase shifts. However, most of the existing ISAC architectures in \cite{lyu2022joint,xiu2024improving,xiu2025secure} consider the UAV as an aerial BS that directly performs both communication and sensing tasks, while the potential of UAVs to assist the BS in sensing remains largely unexplored. 
	
	To address the aforementioned limitation, this paper proposes a practical architecture in which a UAV assists a BS in sensing via controllable AN transmission. 
	While UAV-generated AN enhances the echo from the sensing target and improves the sensing capability of the BS, it also introduces interference through the UAV–BS link and to the communication user. 
	To mitigate the AN-induced interference, a null-space-based deep reinforcement learning (DRL) framework is proposed, in which the UAV beamforming is first projected onto the null space of the UAV–BS channel to suppress the direct interference, and then refined via convex optimization to maximize the sensing accuracy. 
	Meanwhile, the UAV trajectory is optimized using the soft actor–critic (SAC) algorithm to further improve the sensing and communication performance.
	Simulation results show that the proposed method significantly reduces the time-averaged CRB by over 10\%, compared with the ISAC system without UAV assistance. 
	Moreover, compared with the benchmarks of the fixed UAV trajectory with the optimized beamforming and the maximum ratio transmission (MRT)-based beamforming, the proposed method achieves a noticeable improvement in the sensing accuracy.
	
	The remainder of this paper is organized as follows. Section II introduces the system model. Section III presents the proposed null-space-based DRL solution. Section IV provides simulation results, and Section V concludes the paper.

	\begin{figure}
		\centering
		\includegraphics[width=0.85\linewidth]{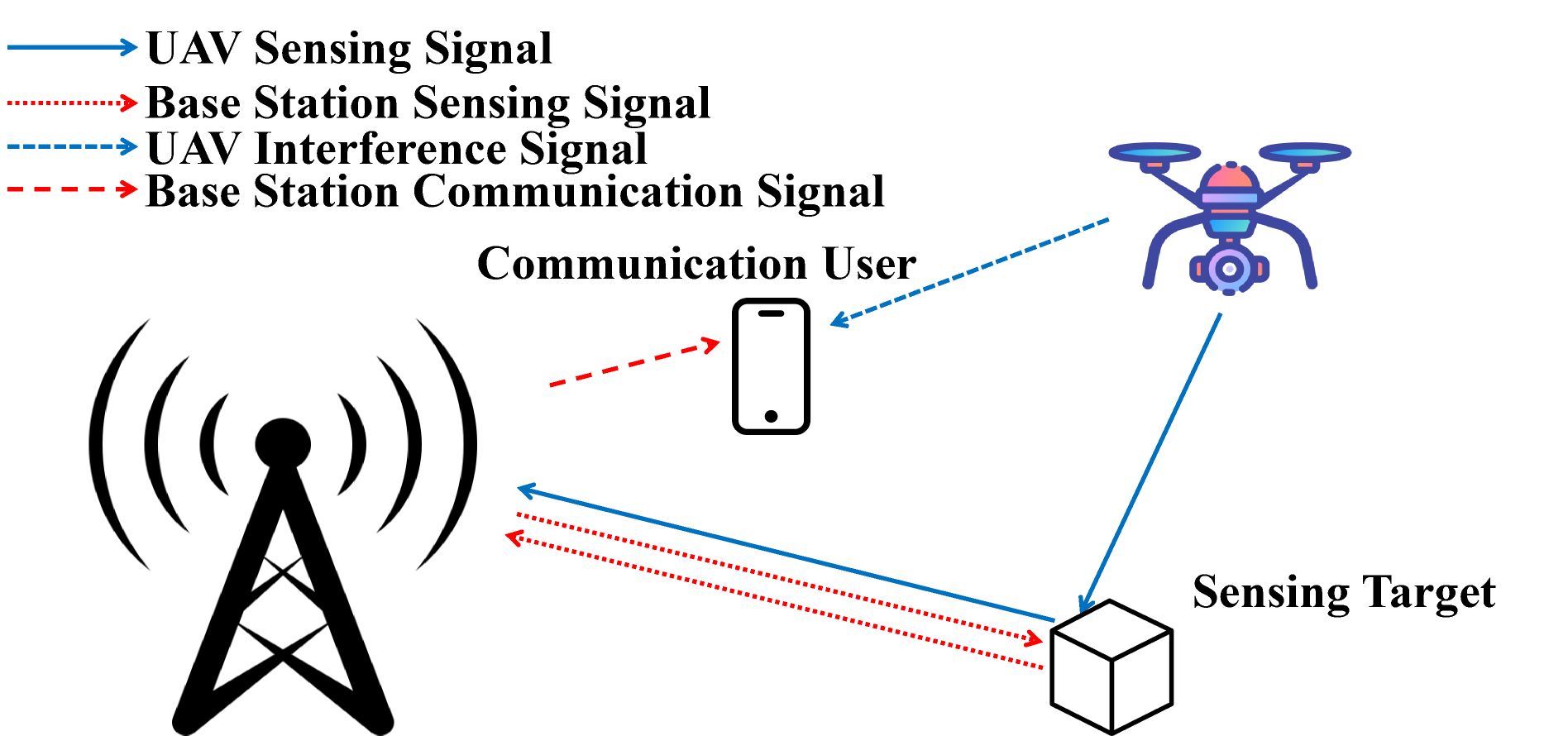}
		\caption{UAV-assisted ISAC system.}
		\label{fig:system} \vspace{-0.5cm}
	\end{figure}

	\section{System Model}
	
	We consider an ISAC system in which a UAV assists a BS in detecting a sensing target, while ensuring that the signal to interference plus noise ratio (SINR) towards a communication user (CU) exceeds a predefined threshold.\footnote{Extensions to multi‑user/multi‑target scenarios will be explored in our future work.}
	As shown in Fig. \ref{fig:system}, all signal links operate over the same frequency band. 
	The UAV is equipped with a uniform linear array (ULA) consisting of $L$ antennas, while the BS employs a ULA with $M$ antennas, and the CU is equipped with a single antenna. 
	The UAV's initial and final positions are denoted by $\boldsymbol q_I$ = [$x_I$, $y_I$, $z_u$] and $\boldsymbol q_F$ = [$x_F$, $y_F$, $z_u$], respectively. 
	It operates at a fixed altitude $z_u$ over a duration $T_u$, with a maximum speed of $V_{max}$. 
	For tractability, the operation period is discretized into $N_{u}=T_{u}/\Delta t$ time slots, with the UAV position   denoted by $\boldsymbol q[n]$ for each time slot $n \in {1, \cdots, N_u}$.

	\subsection{Communication Model}
	
	For downlink transmission, the BS sends the communication signal to the CU as
	\begin{equation}
		\boldsymbol s_{b}[n]= \boldsymbol{w}_{b}[n]x_{b}[n],
	\end{equation}
	where $\boldsymbol{w}_{b}[n] \in \mathbb{C} ^{M \times 1}$ denotes the beamforming vector with transmit power $P_{bm}$, and $x_{b}[n]$ is the unit-power symbol.
	Meanwhile, the UAV transmits the AN
	\begin{equation}
		\boldsymbol s_{u}[n] = \boldsymbol{w}_{u}[n]x_{u}[n],
	\end{equation}
	where $\boldsymbol{w}_{u}[n] \in \mathbb{C} ^{L \times 1}$ is the UAV's beamforming vector,  and $x_{u}[n]\sim \mathcal{CN}(0,1)$ represents the AN signal.  
	The UAV's transmit power is limited by the onboard energy, i.e., $|\boldsymbol{w}_{u}[n]|^2\leq P_{um}$. 
	Therefore, the received signal at the CU is given as\footnote{The reflected signal from the sensing target to the CU is negligible compared with the direct links from the BS and the UAV.} 
	\begin{equation}
		y_{c}[n] = \mathbf H_{bc}[n]\boldsymbol s_{b}[n] + \mathbf H_{uc}[n]\boldsymbol s_{u}[n] + z_c[n],
	\end{equation}
	where $\mathbf{H}_{bc}[n] = \sqrt{\frac{{\beta}_0}{(K_R+1)d_{bc}^2[n]}}(\sqrt{K_R} \mathbf G^H_1[n] + \mathbf G^H_2[n])\in \mathbb{C} ^{1\times M}$ is the Rician channel between the BS and the CU with $K_R$ as the Rician factor, 
	${\beta}_0$ as the channel gain at a reference distance, and
	$d_{bc}[n]$ as the BS-CU distance. 
	Here, $\mathbf G_{1}[n] =  [e^{-j\frac{\pi(M-1)}{2}\sin\theta_{bc}[n]}, e^{-j\frac{\pi(M-3)}{2}\sin\theta_{bc}[n]}, \cdots,$ $e^{j\frac{\pi(M-1)}{2} \sin\theta_{bc}[n]}]^T$ is the steering vector of the BS  with $\theta_{bc}[n]$ as angle-of-departure (AoD) and  ${\mathbf G_2[n]} \sim \mathcal{CN} (\boldsymbol 0,\mathbf I)$ denotes a complex Gaussian random variable. 
	The UAV-CU LoS channel is $\mathbf H_{uc}[n] = \sqrt{{{\beta}}_0}d^{-1}_{uc}[n]\mathbf G_{u}^H[n] \in \mathbb{C} ^{1\times L}$  with $d_{uc}[n]$ as the distance, $\mathbf G_u[n] =  [e^{-j\frac{\pi(L-1)}{2}\sin\theta_{uc}[n]}, e^{-j\frac{\pi(L-3)}{2}\sin\theta_{uc}[n]}, \cdots,$ $ e^{j\frac{\pi(L-1)}{2} \sin\theta_{uc}[n]}]^T$ as the UAV's steering vector, and $\theta_{uc}[n]$ as the AoD of the UAV-CU link. Finally, $z_c[n]\sim \mathcal{CN}(0,\sigma_c^2)$ is the  Gaussian noise. % at the CU. %additive white
	
	Assuming perfect channel state information is available at both the BS and the UAV, the received SINR at the CU in time slot $n$ is given by\footnote{The impact of channel estimation errors caused by mobility will be discussed in our future work.}  
	\begin{equation}
		\tau_c[n] = \frac{|\mathbf H_{bc}[n]\boldsymbol{w}_{b}[n]|^2}
		{\sigma_c^2 + |\mathbf H_{uc}[n] \boldsymbol{w}_{u}[n]|^2}.
	\end{equation}

	\subsection{Sensing Model}
	
	To estimate the location of the sensing target, the BS processes the reflected signal from the target. 
	Meanwhile, the AN transmitted by the UAV aims to enhance the power of the reflected echo.   
	At time $n$, the received echo signal $\boldsymbol y_{s}[n]\in \mathbb{C} ^{M\times 1}$ at the BS can be expressed as\footnote{The UAV beamforming will be designed to ensure zero interference toward the BS, as detailed later in Section II-C.}
	\begin{equation}
		\begin{split}
			\boldsymbol y_{s}[n] = &\alpha_{bs}[n]\boldsymbol a_r (\theta_{bs}[n])\boldsymbol a_t^H(\theta_{bs}[n])\boldsymbol s_b[n]
			\\& + \alpha_{us}[n]\boldsymbol a_r (\theta_{bs}[n])\boldsymbol a_t^H(\theta_{us}[n])\boldsymbol s_{u}[n] + \boldsymbol z_s[n],
		\end{split}
	\end{equation}
	where $\alpha_{bs}[n]$ and $\alpha_{us}[n]$ represent the combined path loss and radar cross section (RCS) coefficients of the BS–target-BS and UAV–target-BS links, respectively. 
	$\theta_{bs}[n]$ and $\theta_{us}[n]$ are the AoD of the sensing target with respect to the BS and the UAV, respectively.
	%$\boldsymbol a_t(\theta_{bs}[n])$ and $\boldsymbol a_r(\theta_{bs}[n])$ are  with 
	$\boldsymbol a_t(\theta_{bs}[n]) = \boldsymbol a_r(\theta_{bs}[n]) = [e^{-j\frac{\pi(M-1)}{2}\sin\theta_{bs}[n]},$ $ e^{-j\frac{\pi(M-3)}{2}\sin\theta_{bs}[n]}, \cdots, e^{j\frac{\pi(M-1)}{2} \sin\theta_{bs}[n]}]^T$ are the transmit and receiving array vectors at the BS.   
	Furthermore, the UAV's transmit array  vector is  $ \boldsymbol a_t(\theta_{us}[n]) = [e^{-j\frac{\pi(L-1)}{2}\sin\theta_{us}[n]},$ $ e^{-j\frac{\pi(L-3)}{2}\sin\theta_{us}[n]}, \cdots,  e^{j\frac{\pi(L-1)}{2} \sin\theta_{us}[n]}]^T$ and $\boldsymbol z_s[n]\sim \mathcal{CN}(\mathbf 0, \sigma_s^2\mathbf {I})$ is the received Gaussian noise. 
	
	The goal of the BS  is to accurately estimate the AoA $\theta_{bs}$ (same as AoD) towards the sensing target, which is critical for target localization. 
	To evaluate the estimation performance,  CRB is adopted to characterize the fundamental lower bound on the variance of unbiased AoA estimation \cite{su2023sensing}. 
	In particular, a smaller CRB implies higher estimation accuracy.
	For notational simplicity, the CRB associated with estimating $\theta_{bs}[n]$ is denoted as $\mathrm{CRB}_{bs}[n]$, as given in the following proposition. 
	\begin{proposition}
		The CRB for estimating $\theta_{bs}[n]$  is
		\begin{equation}
			\begin{split}
				\mathrm{CRB}_{bs}[n] = & [\frac{2}{\sigma_s^2}(|\alpha_{bs}[n]|^2|\boldsymbol a^H_t(\theta_{bs}[n])\boldsymbol s_{b}[n]|^2
				\\ & +\frac{\zeta^2[n] M}
				{\sigma_s^2 + \zeta[n] M })\lVert \frac{\partial \boldsymbol a_r (\theta_{bs}[n])}{\partial \theta_{bs}[n]}\rVert^2]^{-1},
			\end{split}
		\end{equation}
		where \mbox{$\zeta[n]=|\alpha_{us}[n]\boldsymbol a_t^H(\theta_{us}[n])\boldsymbol s_{u}[n]|^2$}.
		
		$\mathnormal{Proof:}$ See Appendix A.
	\end{proposition}

	From Proposition 1, the value of $\mathrm{CRB}_{bs}[n]$ decreases as $\zeta[n]$ increases, where the value of $\zeta[n]$ is jointly determined by the UAV's position $\boldsymbol q[n]$ and beamforming $\boldsymbol{w}_u[n]$. Specifically, $\boldsymbol q[n]$ affects the reflection coefficient $\alpha_{us}[n]$ and steering vector $\boldsymbol a_t^H(\theta_{us}[n])$, while $\boldsymbol{w}_u[n]$ determines the UAV transmit power. 
	The remaining parameters in (6) are dictated by the BS and the environment, which are not controllable. 
	Therefore, improving the sensing accuracy requires minimizing $\mathrm{CRB}_{bs}[n]$, which is equivalent to maximizing the value of $\zeta[n]$ via UAV control. % needs to be maximized.
	
	\subsection{Problem Fromulation}
	Consequently, the objective of the UAV in the ISAC system is to minimize the time-averaged CRB of $\theta_{bs}$ by jointly optimizing its trajectory $\boldsymbol{q}$ and beamforming vector $\boldsymbol{w}_u$, while ensuring zero interference to the BS and maintaining a minimum SINR towards the CU in each time slot. 
	The resulting optimization problem can be formulated as:
	\begin{subequations}\label{equs_Opt}  
		\begin{align}
			\min_{ \boldsymbol{q}[n], \boldsymbol{w}_u[n] } \quad &      \frac{1}{N_u}\sum_{n=1}^{N_u} \mathrm{CRB}_{bs}[n]  \label{equ_obj}\\ 
			\textrm{s. t.} \quad  
			& \tau_c[n] \geq  \gamma, \label{cons_sinr}\\
			& \boldsymbol{a}_t^H(\theta_{bu}[n])\boldsymbol{w}_{u}[n] =0,  \label{cons_interference}\\  
			& \Vert \boldsymbol{w}_{u}[n] \Vert^2 \leq P_{um}, \label{cons_power}\\
			& \boldsymbol{q}_u[1] = \boldsymbol{q}_I, \boldsymbol{q}_u[N_u] = \boldsymbol{q}_F, \label{cons_position}\\
			& \Vert \boldsymbol{q}_u[n]-\boldsymbol{q}_u[n-1]\Vert \leq V_{max} \Delta t, \label{cons_speed}
		\end{align}
	\end{subequations}
	where (\ref{cons_sinr}) ensures a minimum SINR at the CU, 
		(\ref{cons_interference}) enforces zero interference over the  UAV–BS link with $\theta_{bu}[n]$ denoting the AoD from the UAV to the BS, 
		(\ref{cons_power}) constrains the UAV transmit power, 
		(\ref{cons_position}) specifies the initial and final UAV positions, and (\ref{cons_speed}) limits the UAV velocity. 
		Based on the CRB definition in (6), problem (7) is challenging to solve due to the strong coupling between the UAV position and beamforming, leading to a non-convex formulation. 
		Moreover, the trajectory design involves sequential decision-making, which further increases the problem complexity.

	\section{Proposed Solution}
	
	To solve the non-convex problem (7) over multiple time slots, a two-step approach is proposed, where beamforming is first designed to satisfy constraints (\ref{cons_sinr})-(\ref{cons_power}), followed by UAV trajectory optimization using DRL.

	\subsection{Beamforming Optimization via Null-Space Projection}
	\label{nspb}
	
	To address the stringent constraint in (\ref{cons_interference}), a subproblem is first considered: assume the UAV position $\boldsymbol{q}$ to be fixed during  each time $n$, then the beamforming optimization becomes
		\begin{subequations}\label{equs_Opt1}  
			\begin{align}
				\min_{ \boldsymbol{w}_u[n] } \quad &       \mathrm{CRB}_{bs}[n]  \label{equ_obj1}\\ 
				\textrm{s. t.} \quad  
				& \tau_c[n] \geq  \gamma, \label{cons_sinr1}\\
				& \boldsymbol{a}_t^H(\theta_{bu}[n])\boldsymbol{w}_{u}[n] =0,  \label{cons_interference1}\\  
				& \Vert \boldsymbol{w}_{u}[n] \Vert^2 \leq P_{um} \label{cons_power1} .
			\end{align}
	\end{subequations}
	First, to satisfy the constraint (\ref{cons_interference1}), an orthogonal projection matrix $\boldsymbol{w}_{\bot}[n]$ is constructed onto the subspace orthogonal to the BS steering vector, given by
	\begin{equation}
		\boldsymbol{w}_{\bot}[n]=\mathbf I - \frac{\boldsymbol a_t(\theta_{bu}[n])\boldsymbol a_t^H(\theta_{bu}[n])}{\boldsymbol a_t^H(\theta_{bu}[n])\boldsymbol a_t(\theta_{bu}[n])}.
	\end{equation}
    That is, the null-space projection matrix $\boldsymbol{w}_{\bot}[n]$ can project any vector onto the subspace orthogonal to the UAV-BS steering direction, thereby ensuring (\ref{cons_interference}). 
	Accordingly, the beamforming vector can be rewritten as
	\begin{equation}
		\boldsymbol{w}_{u}[n]= \boldsymbol{w}_{\bot}[n]\boldsymbol{w}_{1}[n],
	\end{equation}
	where $\boldsymbol{w}_{1}[n]$ represents the beamforming weight vector within the projected subspace. 
	The vector $\boldsymbol{w}_{1}[n]$ is then optimized to satisfy the remaining constraints while minimizing the CRB.
		Thus, problem (\ref{equs_Opt1}) can be equivalently reformulated as
		\begin{subequations}\label{equs_Opt2}  
			\begin{align}
				\max_{ \boldsymbol{w}_1[n] } \quad &   |\boldsymbol a_t^H(\theta_{us}[n])\boldsymbol{w}_{\bot}[n]\boldsymbol{w}_{1}[n]|^2     \label{equ_obj2}\\ 
				\textrm{s. t.} \quad  
				& \frac{|\mathbf H_{bc}[n]\boldsymbol{w}_{b}[n]|^2}
				{\sigma_c^2 + |\mathbf H_{uc}[n] \boldsymbol{w}_{\bot}[n]\boldsymbol{w}_{1}[n]|^2} \geq  \gamma, \label{cons_sinr2}\\ 
				& \Vert \boldsymbol{w}_{\bot}[n]\boldsymbol{w}_{1}[n] \Vert^2 \leq P_{um} \label{cons_power2}. 
			\end{align}
	\end{subequations}
	Note that problem (\ref{equs_Opt2})  is a non-convex quadratically constrained quadratic program. To facilitate tractable optimization, a standard semidefinite relaxation (SDR) approach is applied by lifting the beamforming vector to a matrix variable. 
		Specifically, define $\mathbf W_1[n] = \boldsymbol{w}_{1}[n]\boldsymbol{w}^H_{1}[n]$, along with 
		$\mathbf A[n] = $ $\boldsymbol{w}^H_{\bot}[n] \boldsymbol{a}_t(\theta_{us}[n]) \boldsymbol{a}_t^H(\theta_{us}[n])\boldsymbol{w}_{\bot}[n] $, 
		$\mathbf B[n] = \boldsymbol{w}^H_{\bot}[n] $ $ \mathbf H^H_{uc}[n] \mathbf H_{uc}[n] \boldsymbol{w}_{\bot}[n]$,  
		and $\mathbf C[n] = \boldsymbol{w}^H_{\bot}[n] \boldsymbol{w}_{\bot}[n]$.  
		Then, problem (\ref{equs_Opt2}) can be equivalently reformulated as
		\begin{subequations}\label{equs_Opt3}  
			\begin{align}
				\max_{\mathbf{W}_1[n] } \quad &   \mathrm{Tr}(\mathbf{A}[n]\mathbf{W}_1[n])     \label{equ_obj3}\\ 
				\textrm{s. t.} \quad  
				& \gamma(\sigma_c^2+\mathrm{Tr}(\mathbf{B}[n]\mathbf{W}_1[n])) \leq  |\textbf H_{bc}[n]\boldsymbol w_{b}[n]|^2, \label{cons_sinr3}\\ 
				& \mathrm{Tr}(\mathbf{C}[n]\mathbf{W}_1[n]) \leq P_{um} \label{cons_power3}, \\ 
				& \mathrm{rank}(\mathbf{W}_1[n])=1, \mathbf{W}_1[n] \succeq 0 \label{cons_rank3}, 
			\end{align}
		\end{subequations}
		where (\ref{cons_rank3}) enforces $\mathbf{W}_1[n]$ to be a rank-one positive semi-definite matrix, ensuring equivalence to the original vector formulation. 
		To address this problem, SDR is applied by relaxing the rank constraint, which transforms the original non-convex problem into a convex semidefinite program that can be efficiently solved. The resulting relaxed optimization problem is given by 
		\begin{subequations}\label{equs_Opt4}  
			\begin{align}
				\max_{\mathbf{W}_1[n] } \quad &   \mathrm{Tr}(\mathbf{A}[n]\mathbf{W}_1[n])     \label{equ_obj4}\\ 
				\textrm{s. t.} \quad  
				& \gamma(\sigma_c^2+\mathrm{Tr}(\mathbf{B}[n]\mathbf{W}_1[n])) \leq  |\mathbf H_{bc}[n]\boldsymbol w_{b}[n]|^2, \label{cons_sinr4}\\ 
				& \mathrm{Tr}(\mathbf{C}[n]\mathbf{W}_1[n]) \leq P_{um} \label{cons_power4}, \\ 
				& \mathbf{W}_1[n] \succeq 0 \label{cons_rank4}. 
			\end{align}
	\end{subequations}
	Note that problem (\ref{equs_Opt4}) is a standard SDP and can be efficiently solved by CVX. However, the obtained solution $\mathbf{W}^{+}_1[n]$ is not guaranteed to be rank one. 
	To recover a feasible solution to the original problem,  Gaussian randomization \cite{luo2010semidefinite} is applied to construct a high-quality rank-one approximation for (\ref{equs_Opt3}). 
	Specifically, $N_{gr}$ candidate vectors will be generated from a complex Gaussian distribution with covariance matrix $\mathbf{W}^{+}_1[n]$, and the candidate that achieves the maximum objective value while satisfying all constraints is selected. 
	The beamforming optimization procedure is summarized in Algorithm 1.  
	The resulting solution $\boldsymbol{w}^*_u[n]$ ensures that, for a given location, the UAV beamforming minimizes the CRB while satisfying all constraints in (\ref{equs_Opt1}). 
	Consequently, the original joint trajectory and beamforming design in (\ref{equs_Opt}) reduces to the trajectory optimization, which is addressed using reinforcement learning as next. 
	
	\begin{algorithm}
		\caption{Null-Space-Based Beamforming Optimization}
		\begin{algorithmic}[1]
			\STATE \textbf{Input:} $\textbf A[n]$, $\textbf B[n]$, $\textbf C[n]$, $\boldsymbol{w}_{\bot}[n]$, and the number of Gaussian random samples $N_{gr}$.
			\STATE Get the  solution $\mathbf{W}^{+}_1[n]$ of problem (\ref{equs_Opt4}) through CVX;
			\IF{$\mathrm{rank}(\mathbf{W}^{+}_1[n]) = 1$}
			\STATE Perform eigenvalue decomposition of $\mathbf{W}^{+}_1[n]$ and recover $\boldsymbol{w}^*_1[n]$ from its principal eigenvector;
			\STATE Get the optimized beamforming vector $\boldsymbol{w}_u^*[n]$ via (10);
			\ELSE
			\STATE Perform eigenvalue decomposition of $\mathbf{W}^{+}_1[n]$ to get eigenvector matrix $\mathbf {U}$ and eigenvalue matrix $\mathbf {\Lambda}$;
			\STATE Generate $N_{gr}$ complex Gaussian random vectors $\boldsymbol {\tilde{v}}_l \sim \mathcal{CN} (\textbf 0,\textbf I), l= 1, \cdots, N_{gr}$;
			\STATE Get $N_{gr}$ candidate solutions $\boldsymbol {\tilde{w}}_l = \mathbf {U}\mathbf {\Lambda}^{\frac{1}{2}}\boldsymbol {\tilde{v}}_l$;
			\STATE Select the best solution $\boldsymbol{w}^*_1[n]$ from all candidates that maximizes (\ref{equ_obj3}) while satisfying  (\ref{cons_sinr3}) and (\ref{cons_power3});
			\STATE Get the optimized beamforming vector $\boldsymbol{w}_u^*[n]$ via (10);
			\ENDIF
			\STATE \textbf{Output:} $\boldsymbol{w}^*_u[n]$.
		\end{algorithmic}
	\end{algorithm}

	\subsection{SAC-based Trajectory Optimization}
	After the beamforming optimization, the trajectory design can be considered as a Markov decision process (MDP) defined by a tuple  $<\mathcal{S}, \mathcal{A}, \mathcal{R}>$, where $\mathcal{S}$ is the state space, $\mathcal{A}$ is the action space, and $\mathcal{R}$ is the reward function. 
	At each time slot $n$, the UAV observes the current state $s[n]$ and selects the action $a[n]$ according to its policy $\pi(a|s)$. 
	The environment then transitions to a new state $s[n+1]$ and returns a reward $r[n]$, which evaluates the action. The  MDP formulation is defined as follows:

	\noindent\textbf{State space}: The state at time slot $n$ is defined as ${s}[n] =  \{x[n], y[n], d[n], \log(\mathrm {CRB}_{bs}[n])\}$, including the current UAV position, the relative distance to the destination $d[n]= \Vert \boldsymbol{q}_u[n]-\boldsymbol{q}_F]\Vert$, and the log transformation of the CRB to improve the training stability.

	\noindent \textbf{Action space}: Given the current state $s[n]$, the UAV selects its action ${a}[n]  = \{v_x[n],v_y[n]\}$, where $v_x[n]$ and $v_y[n]$ denote the UAV velocities along the $x$- and $y$-axes, respectively.

	\noindent \textbf{Reward}: To minimize the CRB while satisfying the constraints, the reward at time slot $n$ is defined as 
	\begin{equation}
		r[n] = R_{d}[n]+R_{s}[n]+R_{ti}[n]+R_{g}[n],
	\end{equation}
	where $R_{d}[n] =\iota_d(d[n-1] - d[n])$ encourages progress toward the destination and  $R_{s}[n] = \iota_s(\log(\mathrm {CRB}_{bs}[n-1])-\log(\mathrm {CRB}_{bs}[n])) $ promotes movement toward regions with lower CRB. 
	Here, $\iota_d > 0$ and $\iota_s >0$ are weighting factors.
	The time penalty $R_{ti}[n]$ is a constant negative number and $R_{g}[n]$ denotes a terminal reward, which is positive if the UAV reaches the destination and zero otherwise.  
	
	Since the UAV control variables lie in a continuous action space, the soft actor–critic (SAC) algorithm with the off-policy DRL framework is adopted.  
	Unlike traditional reinforcement learning methods that only aim to maximize cumulative rewards, SAC also maximizes policy entropy to encourage exploration and mitigate convergence to a local optimal \cite{haarnoja2018soft}. 
	The optimal policy function $\mathbf{\pi}^*$ of SAC is defined as
	\begin{equation}
		\mathbf{\pi}^* = \arg \max_{\pi} \mathbb{E}_{\pi}[\sum_n r({s}[n],{a}[n])+\kappa \mathcal{H}(\pi(\cdot|s[n]))],
	\end{equation}
	where $\mathcal{H}(\pi(\cdot|{s}[n]))$ represents the policy entropy and $\kappa$ is the temperature parameter. The loss function of $\kappa$ can be expressed as $J(\kappa)= \mathbb{E}_{s[n] \sim D, a[n] \sim \pi(\cdot|s[n])}$ $[-\kappa \log \pi(a[n]|s[n])-\kappa \mathcal{H}_{0}]$, where $D$ represents the replay buffer and $\mathcal{H}_{0}$ is desired minimum expected entropy.
	
	\begin{figure}
		\centering
		\vspace{-0.2cm}
		\includegraphics[width=0.8\linewidth]{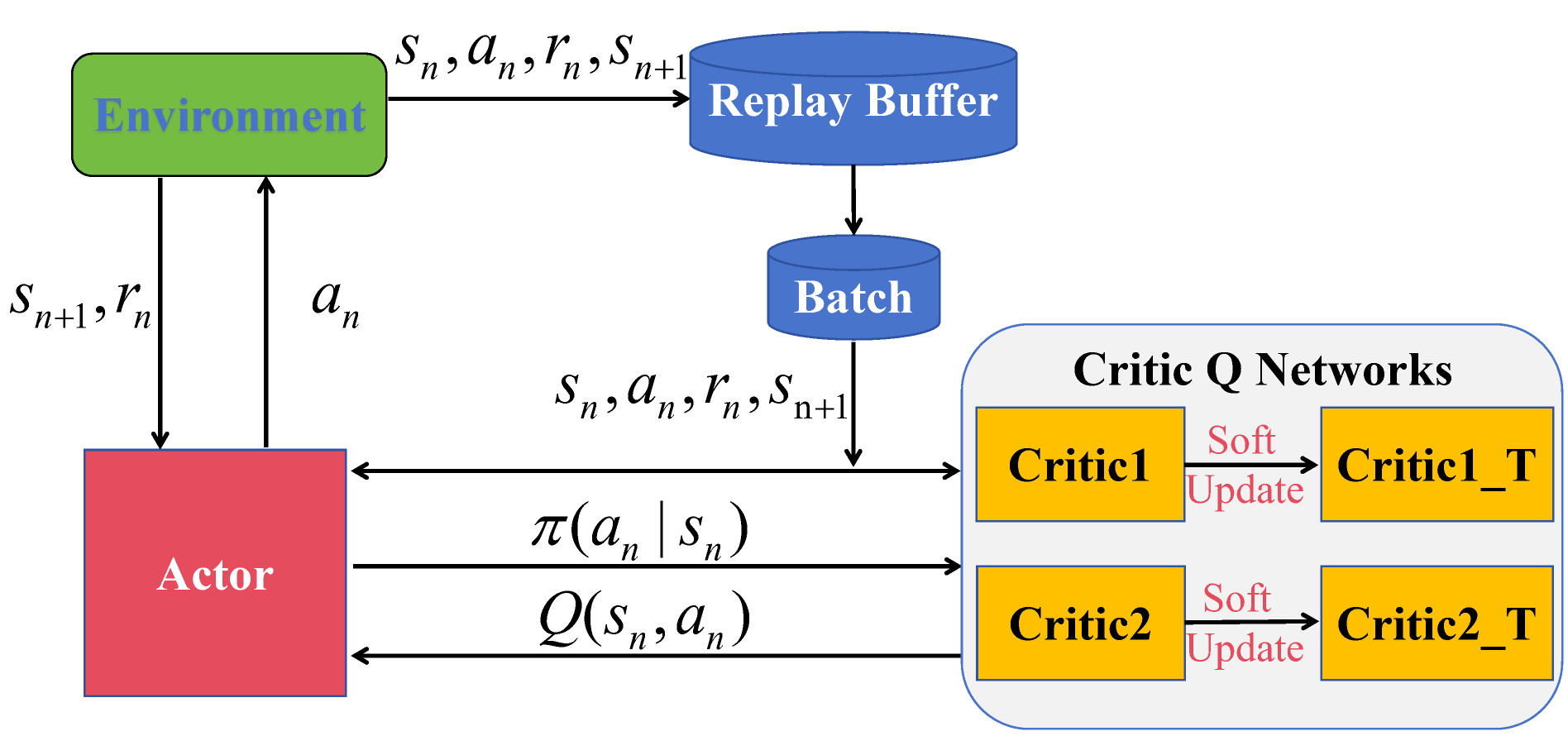}
		\vspace{-0.3cm}
		\caption{Framework of the soft Actor-Critic algorithm.}
		\vspace{-0.5cm}
		\label{fig:SAC}
	\end{figure}
	
	As shown in Fig. \ref{fig:SAC}, the SAC architecture consists of an actor network $\pi(\phi)$ and two critic  networks $Q_{\vartheta_1}$ and $Q_{\vartheta_2}$ with their corresponding target networks $Q_{\overline{\vartheta}_1}$ and $Q_{\overline{\vartheta}_2}$. 
	The actor network generates the policy which defines the distribution of actions given the state, while the critic networks evaluate the action and improve the policy. 
	To mitigate the overestimation of Q-values, the minimum Q-value obtained from the two critic networks is used for policy evaluation. 
	Accordingly, the action-value function $Q(s[n], a[n])$ and the state-value function $V(s[n])$ can be expressed as:
	\begin{equation}
		V(s[n]) = \mathbb{E}_{a[n] \sim \pi}[Q(s[n],a[n])-\kappa \log \pi(a[n]|s[n])],
	\end{equation}
	\begin{equation}
		Q(s[n],a[n]) = r(s[n],a[n])+\chi \mathbb{E}_{s[n+1]}[V(s[n+1])],
	\end{equation}
	where $\chi$ represents the discount coefficient. Thus, the temporal difference loss function of the Q network $J_Q(\vartheta)$ is given as:
	\begin{equation}
		\begin{split}
			J_Q(\vartheta) = &\mathbb{E}_{(s[n],a[n],r[n],s[n+1])\sim D}[(Q(s[n],a[n])
			\\&-(r[n]+\chi(\min \limits_{i=1,2}Q_{\vartheta_{i}}(s[n+1],a[n+1])
			\\&-\kappa \log \pi(a[n+1]|s[n+1]))))^2].
		\end{split}
	\end{equation}
	Based on the critic-estimated Q-values, the actor network updates its parameters by minimizing the Kullback-Leibler (KL) divergence. The loss function of the actor network $J(\phi)$ is given as:
	\begin{equation}
		\begin{split}
			J(\phi) = &\mathbb{E}_{s[n]\sim D, a[n] \sim \pi_\phi}[\kappa \log (\pi_\phi(a[n]|s[n]))\\&- \min \limits_{i=1,2} Q_{\vartheta_{i}}(s[n],a[n])].
		\end{split}
	\end{equation}
	
	Algorithm 2 summarizes the proposed null-space-based DRL framework for UAV-assisted ISAC system optimization. 
	The complexity analysis of the proposed solution is given as follows.
	For Algorithm 1, given a solution accuracy $\varepsilon$, the computational complexity using the interior point method is $\mathcal{O}(\log (\frac{1}{\varepsilon}) L^{6.5})$ \cite{karipidis2008quality}. 
	For the SAC solution part, let $N$ denote the number of layers in the networks, $\psi$ denote the number of nodes in each layer, and $B$ denote the batch size in the training.
	Therefore, the complexity of the SAC framework is $\mathcal{O}(6BN\psi^2)$. 
	Consequently, the computational complexity of the proposed solution in Algorithm 2 is $\mathcal{O}(\log (\frac{1}{\varepsilon}) L^{6.5}+6BN\psi^2)$.

	\begin{algorithm}
		\caption{Null-Space-based DRL algorithm}
		\begin{algorithmic}[1]
			\STATE Initialize the environment, replay buffer \(D\),  batch size $B$, and  parameters $\vartheta_1$, $\vartheta_2$, $\phi$.
			\FOR{each episode step}
			\FOR{each environment step}
			\STATE Get the state $s[n]$ of the UAV;
			\STATE Calculate the beamforming vector via Algorithm 1;
			\STATE Select the action through $\pi_\phi(a[n]|s[n])$;
			\STATE Perform the action and get reward $r[n]$;
			\STATE Update the environment and get next state $s[n+1]$;
			\STATE Store  $<s[n],a[n],r[n],s[n+1]>$ in replay buffer.
			\ENDFOR
			\FOR{each gradient step}
			\STATE Update the critic networks with (18);
			\STATE Update the actor network with (19);
			\STATE Update the temperature parameter;
			\STATE Soft update the target critic networks.
			\ENDFOR
			
			\ENDFOR
			
		\end{algorithmic}
	\end{algorithm}
	
	\section{Simulation Results}
	
	\begin{table}
		\centering
		\caption{\label{tab:widgets}\textsc{Simulation Parameters}.}
		\begin{tabular}{@{}l|l|l|l@{}}
			Parameters & Values & Hyperarameters & Values \\\hline
			Num of BS antennas $M$  & 8 & Num of layers $N$& 3 \\
			Num of UAV antennas $L$  & 8 & Hidden units per layer  & 128\\ 
			Target RCS $\epsilon_{rcs}$  & 4 $\mathrm{m}^2$ & Activation function & ReLU \\
			Signal wavelength $\lambda$   & 0.125 m & Optimizer  & Adam\\
			Antenna gain $G_T$ ($G_R$)  & 8 dBi & Batch size $B$ & 128\\
			Noise power $\sigma_s^2$ ($\sigma_c^2$)   & -40 dBm & Loss  & MSE\\
			BS transmit power $P_{bm}$  & 48 dBm & Discount factor $\chi$ & 0.99  \\
			UAV transmit power $P_{um}$   & 40 dBm & Soft update rate  & 0.005\\
			Reference channel gain ${\beta}_0$   & -30 dB\\
			Rician factor $K_R$ & 9
		\end{tabular}
		\vspace{-0.2cm}
	\end{table}

	\begin{figure}
		\centering
		\includegraphics[width=0.85\linewidth]{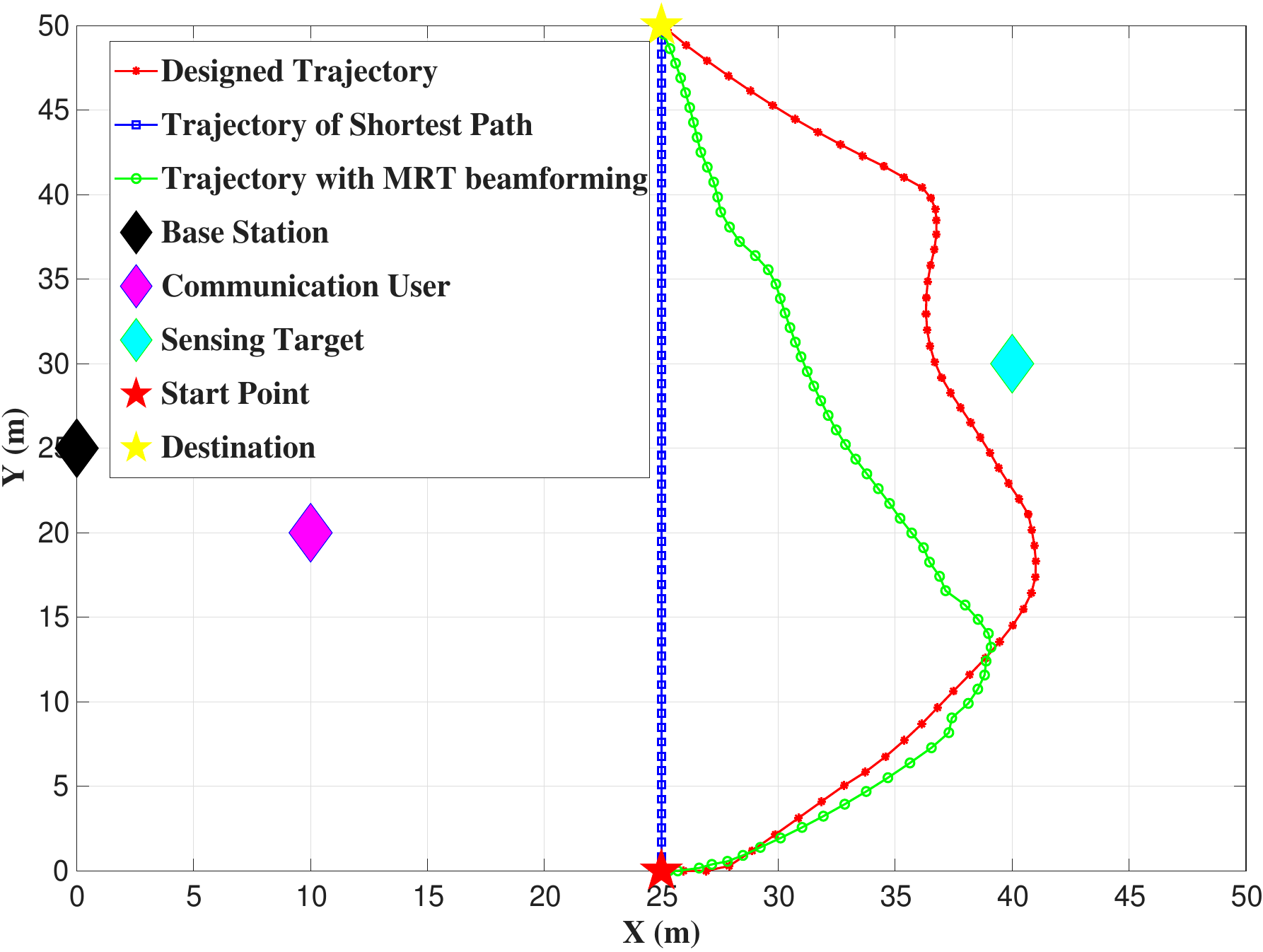}
		\caption{Trajectories of the proposed method and benchmarks.}
		\vspace{-0.5cm}
		\label{fig:3}
	\end{figure}
	
	\begin{figure}
		\centering
		\includegraphics[width=0.85\linewidth]{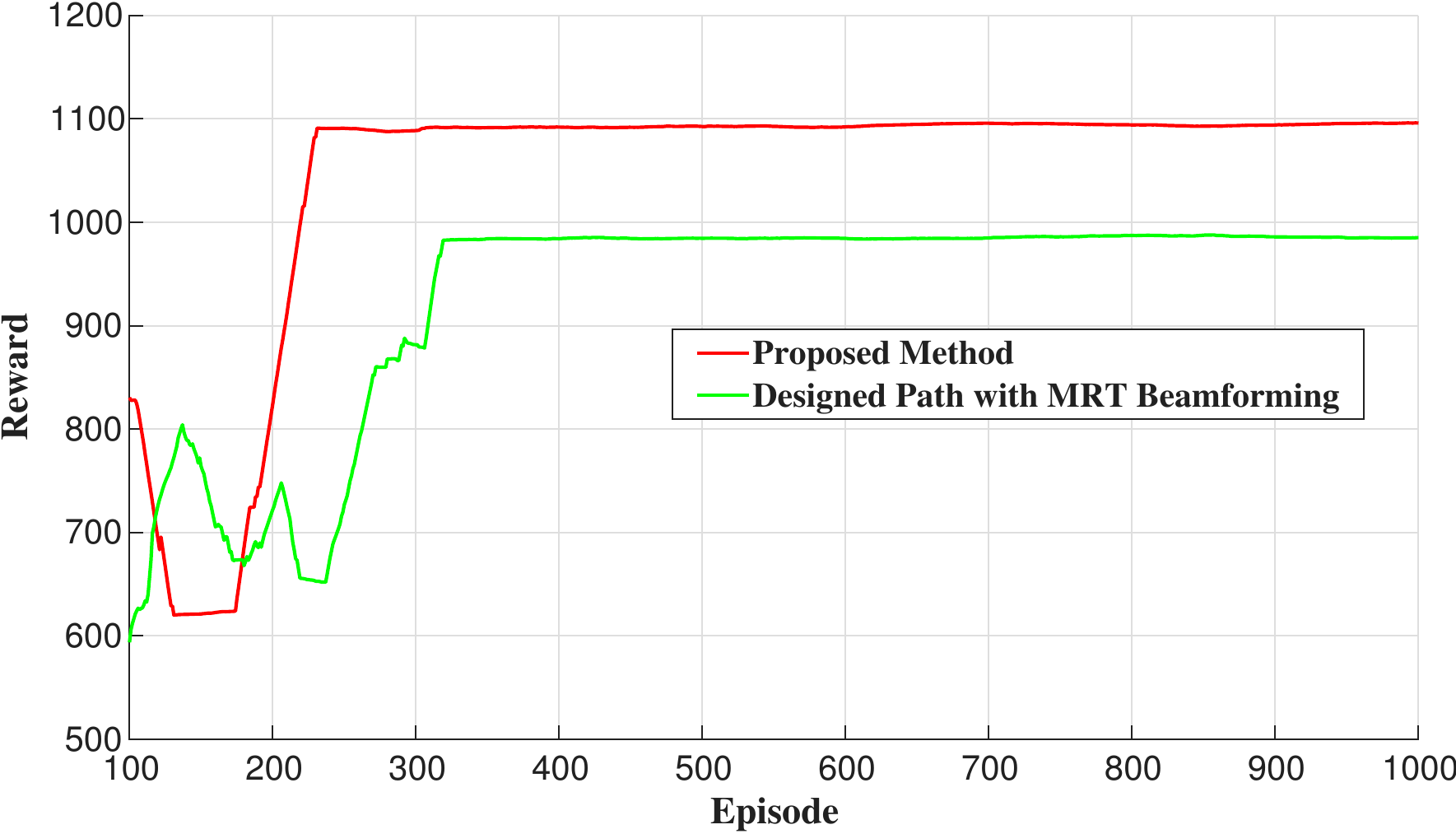}
		\caption{Convergence performance of the proposed method and the MRT-beamforming benchmark with trajectory design.}
		\vspace{-0.5cm}
		\label{fig:5}
	\end{figure}
	
	\begin{figure*}[h]
		\centering
		\begin{subfigure}[t]{0.32\textwidth}
			\centering
			\includegraphics[width=\linewidth]{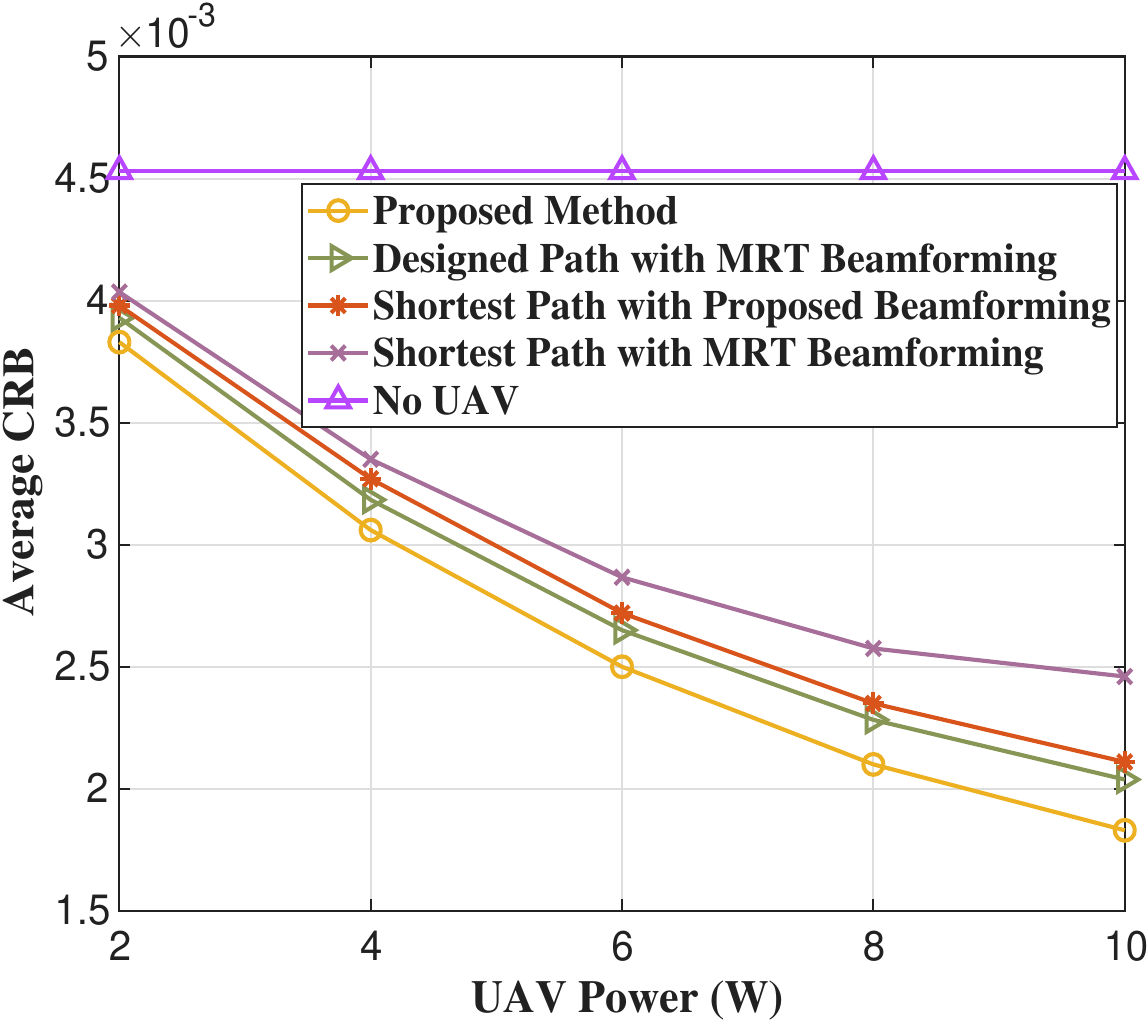}
			\caption{}
			\label{fig:4a}
		\end{subfigure}
		\hfill
		\begin{subfigure}[t]{0.32\textwidth}
			\centering
			\includegraphics[width=\linewidth]{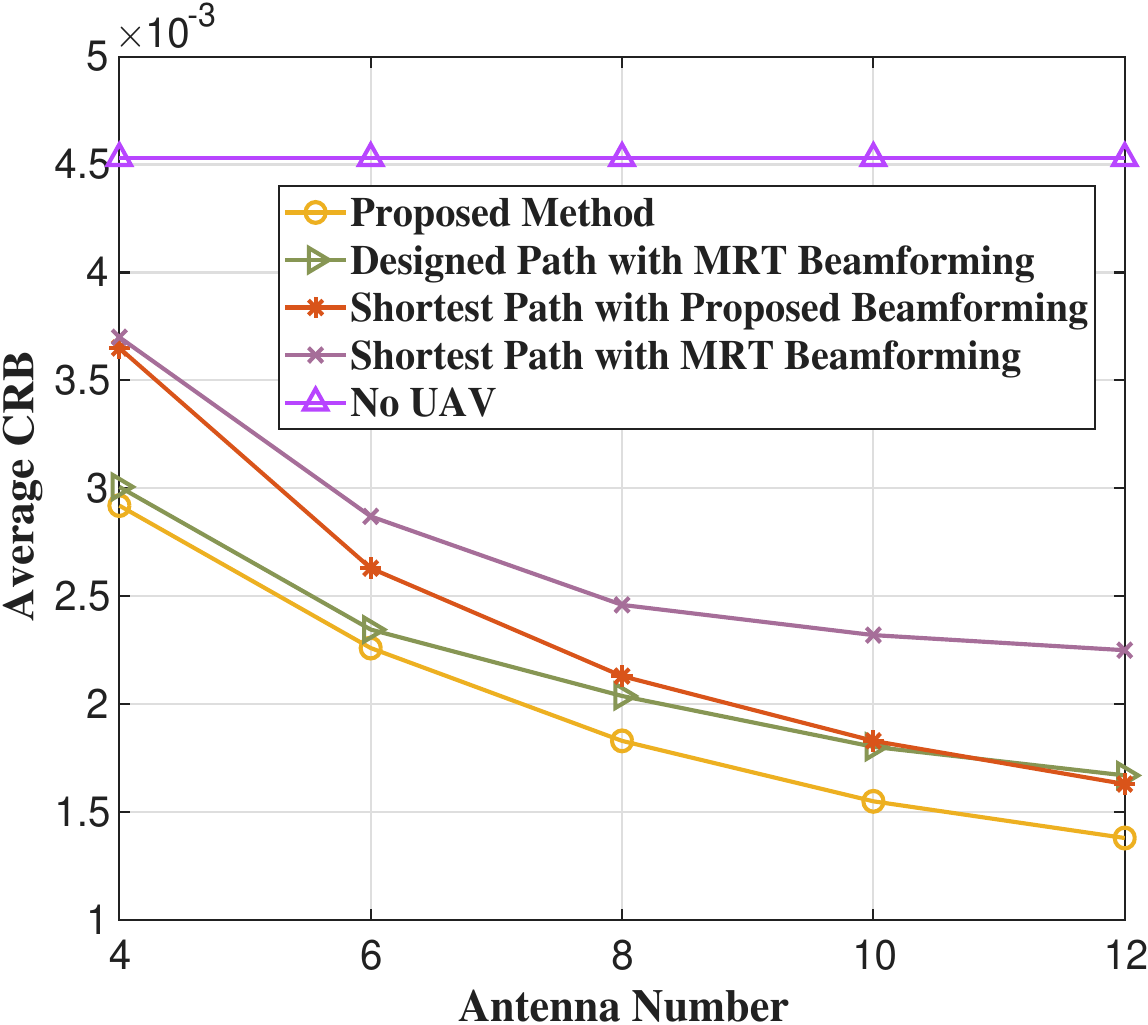}
			\caption{}
			\label{fig:4b}
		\end{subfigure}
		\hfill
		\begin{subfigure}[t]{0.32\textwidth}
			\centering
			\includegraphics[width=\linewidth]{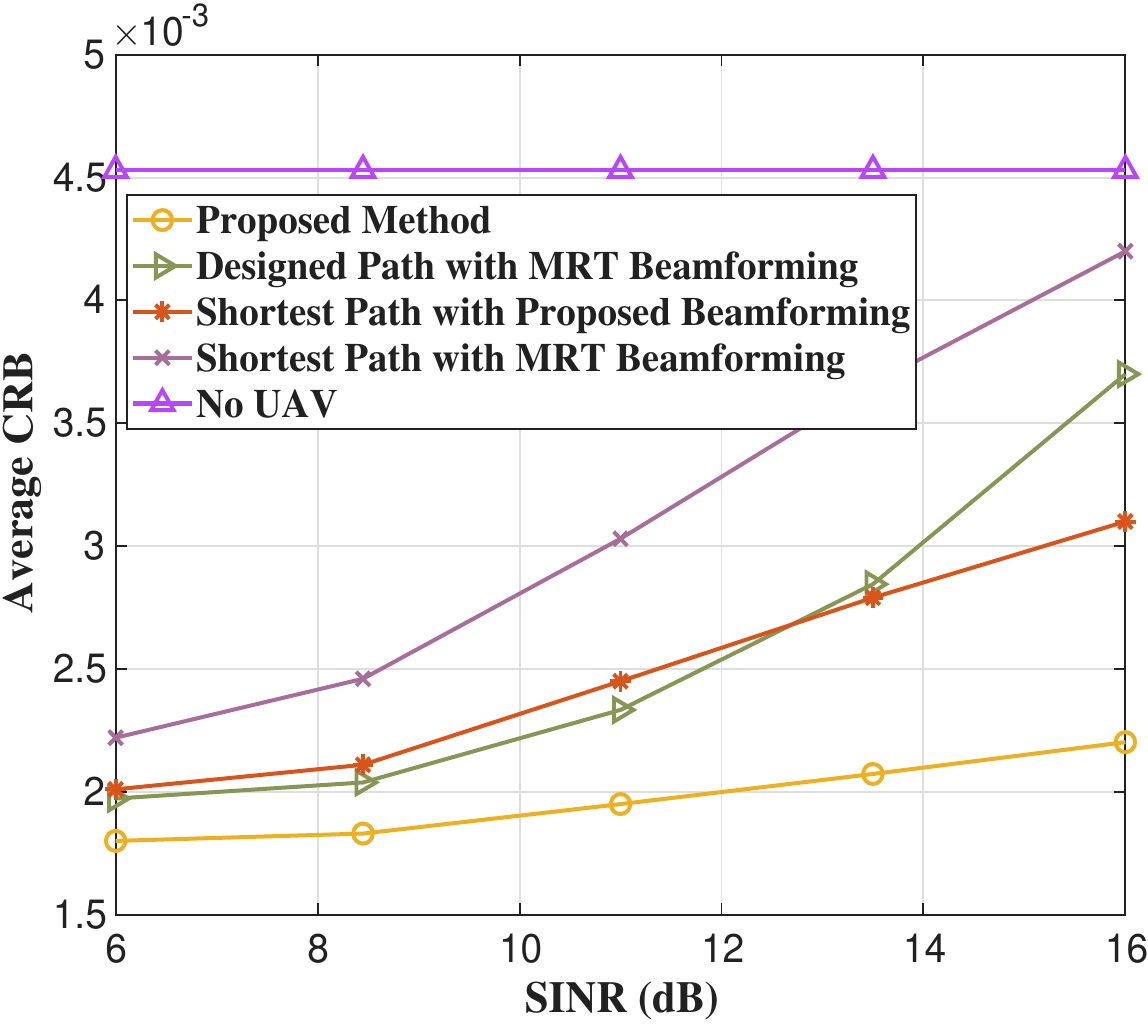}
			\caption{}
			\label{fig:4c}
		\end{subfigure}
		\vspace{-0.2cm}
		\caption{Average CRB vs. (a) UAV maximum transmit power, (b) UAV antenna number, and (c) CU SINR constraint.}
		\vspace{-0.5cm}
		\label{fig:four}
	\end{figure*}

	This section presents simulation results for the proposed UAV-assisted ISAC framework. 
	As shown in Fig. \ref{fig:3}, the BS, CU, and sensing target are located at $(0,25)$, $(10,20)$, and $(40,30)$ meters, respectively, and their positions are fixed throughout the process. The BS and UAV heights are $10$ and $30$ meters, while the CU and target are fixed at $1.5$ meters. The UAV flies from the initial position $(25,0)$ to the destination $(25,50)$, with maximum speeds of 1 m/s along each of the $x$- and $y$-axes.
	The reflection coefficient is $\alpha = \sqrt{\frac{G_TG_R\lambda^2\epsilon_{rcs}}{{(4\pi)}^3d_T^2d_R^2}}$ \cite{yuan2026hybrid}, where $G_T$ and $G_R$ represent the antenna gain of the sensing transmitter and receiver, respectively, $\epsilon_{rcs}$ represents the RCS of the sensing target, $d_T$ represents the distance of between the sensing transmitter and target, $d_R$ represents the distance between the sensing receiver and target, and $\lambda$ denotes the signal wavelength. System parameters and SAC hyperparameters are listed in Table I.   
	To evaluate the performance of the proposed method, four benchmark schemes are considered: 1) the ISAC system that operates without UAV assistance; 2) UAV following the shortest path with beamforming optimization through Algorithm 1 but no trajectory design; 3) UAV using the shortest path and the MRT-based beamforming optimization instead of Algorithm 1. The resulting beamforming vector is given by $\boldsymbol{w}_{u}[n]= \sqrt{P_u[n]}\frac{\boldsymbol{w}_{\bot}[n]\boldsymbol{a}_t(\theta_{us}[n])}{\lVert\boldsymbol{w}_{\bot}[n]\boldsymbol{a}_t(\theta_{us}[n])\rVert}$, where the beamforming direction is designed and fixed to maximize the transmit gain toward the sensing target, and only the transmit power $P_u[n]$ is optimized to minimize the CRB. 4) In the fourth benchmark, the beamforming direction is designed according to the MRT criterion, and the DRL algorithm will jointly optimize the UAV position and transmit power $P_u[n]$, yielding an enlarged state and action space and increasing the coupling between the trajectory and power control.
	
	The resulting trajectories of the proposed method, the shortest-path benchmark, and the benchmark 4 (designed trajectory with MRT beamforming) are shown in Fig. \ref{fig:3}. The difference between the proposed method and the benchmark 4 is mainly caused by the fact that the MRT-based approach leaves the power optimization to the learning algorithm, thus enlarging the state-action space as well as the optimization complexity. Consequently, the learning process becomes more challenging, leading the UAV to adopt a more conservative trajectory and turn toward the destination at an earlier stage.  Fig. \ref{fig:5} shows the convergence performance with the reward learning curve of the proposed method and the MRT-beamforming benchmark with trajectory design, which is smoothed using the 100-episode moving average. Both curves show that the training reward eventually reaches a steady stage, which verifies the convergence behavior of the learning algorithm. Moreover, Fig. \ref{fig:5} shows that the proposed method requires fewer training episodes to reach the steady stage than the state-of-the-art benchmark, owing to its smaller state and action spaces, as well as the reduced training complexity.
	
	Fig. \ref{fig:4a} illustrates the average CRB performance of the proposed method and benchmarks as the maximum transmit power of the UAV increases from $2$ to $10$ watts. Compared with the ISAC system without UAV assistance, the proposed scheme achieves a consistently lower CRB, demonstrating the effectiveness of UAV-aided sensing. 
	In addition, the average CRB of the optimized trajectory yields a better performance over the shortest-path approaches. Meanwhile, the proposed method outperforms the MRT-based beamforming benchmark, which validates the effectiveness of the proposed optimization of the beamforming direction and power allocation in our solution.
	Furthermore, as the transmission power of the UAV increases, the CRB is monotonically decreasing, indicating that the higher power improves the sensing performance. 
	However, the gain from increased transmit power gradually saturates due to the SINR constraint. As the transmit power increases, the gap between the proposed method and all benchmarks becomes larger. This result highlights the great efficacy of the proposed method in enhancing sensing performance under system constraints.
	
	Next, we evaluate the impact of the UAV antennas number on the average CRB. 
	As shown in Fig. \ref{fig:4b}, the proposed method reduces the average CRB by over 10$\%$, compared with the ISAC without UAV assistance. 
	As the number of antennas increases, the average CRB of our proposed method decreases monotonically due to the increase in array gain. 
	Meanwhile, the proposed method consistently outperforms the shortest-path trajectory, indicating that trajectory optimization enables more effective utilization of the array gain. In addition, as the array gain increases, the CRB of the MRT-based beamforming benchmark saturates earlier than the proposed method. This is mainly because the MRT-based approach does not optimize the beamforming direction as efficiently as our proposed solution in Algorithm 1, making the benchmark more susceptible to the CU SINR constraint. Therefore, the UAV has to reduce the transmit power to satisfy the communication constraint, thereby sacrificing the sensing performance.

	Finally, Fig. \ref{fig:4c} presents the variation of the CRB with the CU's SINR constraint. The average CRB increases as the SINR constraint increases, which shows the trade-off between sensing and communication performance. 
	For a larger SINR constraint, the UAV needs to restrict the transmit power to limit the interference to the CU, which in turn reduces the sensing capability of the system. 
	Besides, compared with the performance of the shortest-path trajectory, the proposed method is less sensitive to the SINR constraint. 
	This is because the proposed method jointly optimizes the trajectory and beamforming, providing additional flexibility to reduce the interference to CU. 
	In addition, the proposed method achieves a slower CRB increase than the MRT-based benchmark as its beamforming vector is fully optimized with no limit to the fixed MRT direction. This additional flexibility enables more effective beamforming, transmit power allocation, and trajectory adaptation to better balance the sensing enhancement and communication constraint.

	\section{Conclusion}
	In this paper, a novel UAV-assisted ISAC system is presented. By adopting the average CRB as the performance metric of sensing, the trajectory and transmit beamforming of the UAV are jointly optimized to minimize the average CRB while satisfying the communication constraints. To solve this problem, a framework that integrates null-space projection-based beamforming design with trajectory optimization via DRL is proposed. Simulation results demonstrate that the proposed method can effectively achieve a higher sensing accuracy while ensuring a satisfied communication performance, compared with multiple state-of-the-art benchmarks. 
	
	\section*{Appendix A}
	
	For notational simplicity, the time index $n$ is omitted in the following derivations. 
	To derive the CRB, we first derive the Fisher information matrix (FIM) associated with the target angle estimation \cite{besson2013fisher}, and then compute the CRB accordingly. 
	Let $\mathbf \xi = [\theta_{bs}, \mathrm{Re}\{\alpha_{bs}\}, \mathrm{Im}\{\alpha_{bs}\}]^T \in \mathbb{R}^{3\times1}$ denote the vector of unknown parameters, where $\theta_{bs}$ is the parameter of interest. 
	Since the reflection coefficient $\alpha_{bs}$ is also unknown, it is treated as a nuisance parameter in the FIM. 
	For notational convenience, we define ${\mathbf \varrho} = [\mathrm {Re}\{\alpha_{bs}\}, \mathrm {Im}\{\alpha_{bs}\}]$. The FIM for estimate $\mathbf{\xi}$ is given by 
	\begin{equation}
		\mathit{\mathbf J} =  \begin{bmatrix}  J_{\theta_{bs},\theta_{bs}} & \mathbf J_{\theta_{bs},\mathbf{\varrho}} \\ \mathbf J_{\theta_{bs},\mathbf{\varrho}}^T & \mathbf J_{\mathbf{\varrho},\mathbf{\varrho}}  \end{bmatrix}.
	\end{equation}
	Based on (5), we can rewrite the received signal at the BS as $\boldsymbol y_{s} \sim \mathcal{CN}(\boldsymbol{\mu}(\theta_{bs}), \mathbf {C}(\theta_{bs}))$, where the mean vector is $\boldsymbol{\mu}(\theta_{bs})= \alpha_{bs}\boldsymbol a_r (\theta_{bs})\boldsymbol a_t^H(\theta_{bs})\boldsymbol s_{b}$ and the covariance matrix is $\mathbf {C}(\theta_{bs}) = [\alpha_{us}\boldsymbol a_r (\theta_{bs})\boldsymbol a_t^H(\theta_{us})\boldsymbol s_{u}][\alpha_{us}\boldsymbol a_r (\theta_{bs})\boldsymbol a_t^H(\theta_{us})\boldsymbol s_{u}]^{H} + \sigma_s^2\textbf I$. 
	According to \cite{besson2013fisher}, the Gaussian expression of FIM is
	\begin{equation}
		\mathbf J_{\mathbf{\theta},\mathbf{\varrho}}=2\mathrm{Re}[\frac{\partial \boldsymbol {\mu}^H}{\partial \theta}\textbf C^{-1}\frac{\partial \boldsymbol {\mu}}{\partial  \varrho}]+\mathrm{Tr}[\textbf C^{-1}\frac{\partial \textbf C}{\partial \theta}\textbf C^{-1} \frac{\partial \textbf C}{\partial \varrho}].
	\end{equation}
	Each element of $\textbf J$ can be calculated as
	\begin{equation}
		\begin{split}
			&J_{\mathbf{\theta}_{bs},\mathbf{\theta}_{bs}}=\frac{2|\alpha_{bs}|^2\lVert\frac{\partial \boldsymbol a_r (\theta_{bs})}{\partial \theta_{bs}}\rVert^2|\boldsymbol a_t^H(\theta_{bs})\boldsymbol s_{b}|^2}{\sigma_s^2}\\&+\frac{2|\alpha_{bs}|^2|\frac{\partial \boldsymbol a_t^H(\theta_{bs})\boldsymbol s_{b}}{\partial \theta_{bs}}|^2M}{\sigma_s^2+\zeta M} + \frac{2}{\sigma_s^2}\frac{\zeta^2M}{\sigma_s^2+\zeta M}\lVert\frac{\partial \boldsymbol a_r(\theta_{bs})}{\partial \theta_{bs}}\rVert^2,
		\end{split}
	\end{equation}
	\begin{equation}
		\begin{split}
			\mathbf J_{\mathbf{\theta}_{bs},\mathbf{\varrho}} = &\frac{2}{\sigma_s^2}\mathrm{Re}\{[\alpha^{*}_{bs}\boldsymbol s_{b}^{H}\frac{\partial \boldsymbol a_t(\theta_{bs})}{\partial \theta_{bs}}M\boldsymbol a^{H}_t(\theta_{bs})\boldsymbol s_{b}\\&-\frac{\zeta M^2 \alpha^{*}_{bs}\boldsymbol s_{b}^{H}\frac{\partial \boldsymbol a_t(\theta_{bs})}{\partial \theta_{bs}}\boldsymbol a^{H}_t(\theta_{bs})\boldsymbol s_{b}}{\sigma_s^2+ \zeta M}][1,j]\},
		\end{split}
	\end{equation}
	\begin{equation}
		\mathbf J_{\mathbf{\varrho},\mathbf{\varrho}} = \frac{2M|\boldsymbol a_t^H(\theta_{bs})\boldsymbol s_{b}|^2}{\sigma_s^2+\zeta M}\textbf I_2.
	\end{equation}
	According to the FIM definition, the CRB can be given by
	\begin{equation}
		\mathrm{CRB}_{bs}=[\textbf J^{-1}]_{11} = [J_{\theta_{bs},\theta_{bs}}-\mathbf J_{\theta_{bs},\mathbf{\varrho}}\mathbf J_{\mathbf{\varrho},\mathbf{\varrho}}^{-1}\mathbf J_{\theta_{bs},\mathbf{\varrho}}^T]^{-1}.
	\end{equation}
	Substituting (22), (23), and (24) in (25), we have
	\begin{equation}
		%\begin{split}
		\mathrm{CRB}_{bs} =  [\frac{2}{\sigma_s^2}(|\alpha_{bs}|^2|\boldsymbol a^H_t(\theta_{bs})\boldsymbol s_{b}|^2+\frac{\zeta^2 M}
		{\sigma_s^2 + \zeta M })\lVert\frac{\partial \boldsymbol a_r(\theta_{bs})}{\partial \theta_{bs}}\rVert^2]^{-1}.
		%\end{split}
	\end{equation}
	
	\nocite{*}
	\bibliographystyle{IEEEtran}
	\bibliography{VTC}

% Generated by IEEEtran.bst, version: 1.14 (2015/08/26)
\begin{thebibliography}{10}
\providecommand{\url}[1]{#1}
\csname url@samestyle\endcsname
\providecommand{\newblock}{\relax}
\providecommand{\bibinfo}[2]{#2}
\providecommand{\BIBentrySTDinterwordspacing}{\spaceskip=0pt\relax}
\providecommand{\BIBentryALTinterwordstretchfactor}{4}
\providecommand{\BIBentryALTinterwordspacing}{\spaceskip=\fontdimen2\font plus
\BIBentryALTinterwordstretchfactor\fontdimen3\font minus
  \fontdimen4\font\relax}
\providecommand{\BIBforeignlanguage}[2]{{%
\expandafter\ifx\csname l@#1\endcsname\relax
\typeout{** WARNING: IEEEtran.bst: No hyphenation pattern has been}%
\typeout{** loaded for the language `#1'. Using the pattern for}%
\typeout{** the default language instead.}%
\else
\language=\csname l@#1\endcsname
\fi
#2}}
\providecommand{\BIBdecl}{\relax}
\BIBdecl

\bibitem{gonzalez2024integrated}
N.~Gonz{\'a}lez-Prelcic, M.~F. Keskin, O.~Kaltiokallio, M.~Valkama, D.~Dardari,
  X.~Shen, Y.~Shen, M.~Bayraktar, and H.~Wymeersch, ``The {I}ntegrated
  {S}ensing and {C}ommunication {R}evolution for 6{G}: {V}ision, {T}echniques,
  and {A}pplications,'' \emph{Proceedings of the IEEE}, vol. 112, no.~7, pp.
  676--723, 2024.

\bibitem{liu2022integrated}
F.~Liu, Y.~Cui, C.~Masouros, J.~Xu, T.~X. Han, Y.~C. Eldar, and S.~Buzzi,
  ``Integrated {S}ensing and {C}ommunications: {T}oward {D}ual-{F}unctional
  {W}ireless {N}etworks for 6{G} and {B}eyond,'' \emph{IEEE Journal on Selected
  Areas in Communications}, vol.~40, no.~6, pp. 1728--1767, 2022.

\bibitem{dong2024sensing}
F.~Dong, F.~Liu, Y.~Cui, S.~Lu, and Y.~Li, ``Sensing as a {S}ervice in 6{G}
  {P}erceptive {M}obile {n}etworks: Architecture, {A}dvances, and the {R}oad
  {A}head,'' \emph{IEEE Network}, vol.~38, no.~2, pp. 87--96, 2024.

\bibitem{zou2024securing}
J.~Zou, C.~Masouros, F.~Liu, and S.~Sun, ``Securing the {S}ensing
  {F}unctionality in {ISAC} {N}etworks: An {A}rtificial {N}oise {D}esign,''
  \emph{IEEE Transactions on Vehicular Technology}, vol.~73, no.~11, pp.
  17\,800--17\,805, 2024.

\bibitem{hou2023secure}
K.~Hou and S.~Zhang, ``Secure {I}ntegrated {S}ensing and {C}ommunication
  {E}xploiting {T}arget {L}ocation {D}istribution,'' in \emph{IEEE Global
  Communications Conference}, Kuala Lumpur, Malaysia, 2023, pp. 4933--4938.

\bibitem{lv2025isac}
L.~Lv, J.~Fu, Y.~Feng, L.~Yang, and A.~Nallanathan, ``{ISAC}-{E}nhanced
  {S}ecrecy {C}ommunication: {A} {T}wo-{S}tage {A}rtificial {N}oise-{A}ssisted
  {S}ecure {B}eamforming {D}esign,'' \emph{IEEE Communications Letters},
  vol.~30, pp. 243--247, 2025.

\bibitem{huang2023unmanned}
N.~Huang, C.~Dou, Y.~Wu, L.~Qian, B.~Lin, and H.~Zhou,
  ``{U}nmanned-{A}erial-{V}ehicle-{A}ided {I}ntegrated {S}ensing and
  {C}omputation with {M}obile-{E}dge {C}omputing,'' \emph{IEEE Internet of
  Things Journal}, vol.~10, no.~19, pp. 16\,830--16\,844, 2023.

\bibitem{lyu2022joint}
Z.~Lyu, G.~Zhu, and J.~Xu, ``{J}oint {T}rajectory and {B}eamforming {D}esign
  for {UAV}-{E}nabled {I}ntegrated {S}ensing and {C}ommunication,'' in
  \emph{IEEE International Conference on Communications}, Seoul, Korea, 2022,
  pp. 1593--1598.

\bibitem{xiu2024improving}
Y.~Xiu, W.~Lyu, P.~L. Yeoh, Y.~Ai, Y.~Li, and N.~Wei, ``{I}mproving
  {P}hysical-{L}ayer {S}ecurity in {ISAC-AAV} {S}ystem: {B}eamforming and
  {T}rajectory {O}ptimization,'' \emph{IEEE Transactions on Vehicular
  Technology}, vol.~74, no.~2, pp. 3503--3508, 2024.

\bibitem{xiu2025secure}
Y.~Xiu, W.~Lyu, P.~L. Yeoh, Y.~Ai, and N.~Wei, ``{S}ecure {E}nhancement for
  {RIS}-{A}ided {UAV} {W}ith {ISAC}: {R}obust {D}esign and {R}esource
  {A}llocation,'' \emph{IEEE Transactions on Vehicular Technology}, vol.~75,
  no.~4, pp. 6776--6791, 2026.

\bibitem{su2023sensing}
N.~Su, F.~Liu, and C.~Masouros, ``{S}ensing-{A}ssisted {E}avesdropper
  {E}stimation: {A}n {ISAC} {B}reakthrough in {P}hysical {L}ayer {S}ecurity,''
  \emph{IEEE Transactions on Wireless Communications}, vol.~23, no.~4, pp.
  3162--3174, 2023.

\bibitem{luo2010semidefinite}
Z.-Q. Luo, W.-K. Ma, A.~M.-C. So, Y.~Ye, and S.~Zhang, ``{S}emidefinite
  {R}elaxation of {Q}uadratic {O}ptimization {P}roblems,'' \emph{IEEE Signal
  Processing Magazine}, vol.~27, no.~3, pp. 20--34, 2010.

\bibitem{haarnoja2018soft}
T.~Haarnoja, A.~Zhou, K.~Hartikainen, G.~Tucker, S.~Ha, J.~Tan, V.~Kumar,
  H.~Zhu, A.~Gupta, P.~Abbeel \emph{et~al.}, ``{S}oft {A}ctor-{C}ritic
  {A}lgorithms and {A}pplications,'' \emph{arXiv preprint arXiv:1812.05905},
  2018.

\bibitem{karipidis2008quality}
E.~Karipidis, N.~D. Sidiropoulos, and Z.-Q. Luo, ``{Q}uality of {S}ervice and
  {M}ax-{M}in {F}air {T}ransmit {B}eamforming to {M}ultiple {C}ochannel
  {M}ulticast {G}roups,'' \emph{IEEE Transactions on Signal Processing},
  vol.~56, no.~3, pp. 1268--1279, 2008.

\bibitem{yuan2026hybrid}
M.~Yuan, D.~He, H.~Yin, H.~Wang, F.~Liu, Z.~Wang, and T.~Q. Quek, ``{H}ybrid
  {B}eamforming for mm{W}ave {I}ntegrated {S}ensing and {C}ommunication with
  {M}ulti-{S}tatic {C}ooperative {L}ocalization,'' \emph{IEEE Transactions on
  Wireless Communications}, vol.~25, pp. 771--786, 2026.

\bibitem{besson2013fisher}
O.~Besson and Y.~I. Abramovich, ``{O}n the {F}isher {I}nformation {M}atrix for
  {M}ultivariate {E}lliptically {C}ontoured {D}istributions,'' \emph{IEEE
  Signal Processing Letters}, vol.~20, no.~11, pp. 1130--1133, 2013.

\end{thebibliography}
	
\end{document}